\documentstyle[12pt,rotate]{article}

\input epsf.sty

\newcommand{\insertfig}[2]{\mbox{\epsfxsize=#1cm \epsfbox{#2.eps}}}

\begin{document}
\begin{titlepage}

{\hfill\parbox{40mm}{{ UMD PP\#01-003\\[2mm]
                       DOE/ER/40762-212}}}

\vspace{2cm}

\centerline{\bf THE SINGLET $g_2$ STRUCTURE FUNCTION}
\centerline{\bf IN THE NEXT-TO-LEADING ORDER}

\vspace{15mm}

\centerline{\bf A.V. Belitsky$^a$, Xiangdong Ji$^b$, Wei Lu$^b$, 
                Jonathan Osborne$^b$}

\vspace{15mm}

\centerline{\it $^a$C.N.\ Yang Institute for Theoretical Physics}
\centerline{\it State University of New York at Stony Brook}
\centerline{\it NY 11794-3840, Stony Brook, USA}

\vspace{0.5cm}

\centerline{\it $^b$Department of Physics, University of Maryland}
\centerline{\it College Park, Maryland 20742 }

\vspace{30mm}

\centerline{\bf Abstract}

\hspace{0.5cm}

Following a previous study of the one-loop factorization of the nonsinglet
$g_2$ structure function of the nucleon, we present in this paper the
next-to-leading order coefficient functions in the singlet sector. To
obtain the result, the partonic processes of virtual Compton scattering
off two and three ``on-shell" gluons are calculated. A key step in
achieving the correct factorization is to separate the correct twist-two
contribution. The Burkardt-Cottingham sum rule is nominally satisfied at
this order.

\end{titlepage}

In lepton-nucleon deep-inelastic scattering with the electromagnetic
current, two spin-de\-pen\-dent structure functions of the nucleon can be
studied: $g_{1,2} (x_B, Q^2)$. $g_1 (x_B, Q^2)$ is closely related to
the spin structure of the nucleon and has been investigated extensively
in the last decade. $g_2 (x_B, Q^2)$ is present in processes involving
transversely polarized nucleons and is a three-twist structure function
in the sense that it contributes to physical observables at order $1/Q$
\cite{hey}. The experimental measurements of the $g_2$ structure function
have been done by several collaborations \cite{exp}. There is a long
history in debating the physics involved in $g_2 (x_B, Q^2)$. As more
results from quantum chromodynamics (QCD) are obtained and understood, it
is generally accepted that $g_2(x_B,Q^2)$ probes quark and gluon
correlations in the nucleon which cannot be accessed through Feynman-type
incoherent parton scattering.

At leading-order, $g_2 (x_B, Q^2)$ can be expressed in terms of a simple
parton distribution $\Delta q_T(x)$,
\begin{equation}
g_T (x_B, Q^2) \equiv ( g_1 + g_2 ) (x_B, Q^2)
= \frac{1}{2} \sum_a e_a^2
\left( \Delta q_{aT} (x_B, Q^2) + \Delta \bar q_{aT} (x_B, Q^2) \right) \ ,
\end{equation}
where
\begin{equation}
\Delta q_T (x) = {1\over 2M} \int {d\lambda\over 2\pi}
{\rm e}^{i\lambda x}
\langle PS_\perp |
\bar \psi(0) \gamma_\perp \gamma_5 \psi (\lambda n)
| PS_\perp \rangle \ .
\end{equation}
This result has lead to many incorrect interpretations of the $g_2$ physics.
When the leading-logarithmic corrections are studied, it is found that
$\Delta q_T(x,Q^2)$ mixes with other more complicated distributions under
scale evolution \cite{Shur}. In fact, $\Delta q_T(x,Q^2)$ is a special
moment of a general class of parton distributions involving two light-cone
variables \cite{bkl,Rat}. When the scale changes, only those general
distributions evolve autonomously.

Thus a general factorization formula for $g_2$ is much more involved than
the leading-order result shows. It should contain the generalized
two-variable distributions, $K_i(x,y)$. Indeed, we shall write in general
\begin{equation}
g_T (x_B, Q^2) = \sum_i \int^1_{-1}
\frac{dx}{x} \frac{dy}{y}
\left\{
C_{i}\left( {x_B\over x}, {x_B\over y}; \alpha_s \right) K_{i} (x, y)
+ ( x_B \rightarrow -x_B)
\right\} \ ,
\label{resuts}
\end{equation}
where $C_i$ are the coefficient functions. In a previous paper
\cite{lastpap}, we studied the factorization of the nonsinglet part at
the one-loop order where there were two distributions, $K_{1a,2a}$,
associated with each nonsinglet component $a$, and their one-loop
coefficient functions were obtained for the first time.

In this paper, we study the one-loop factorization of the singlet part of
$g_T (x_B, Q^2)$. The subject has previously been considered in Ref.\
\cite{Bel}, a comparison of the results will be made in the end of the
paper. Contrary to the previous conclusion, the result of this paper
represents a local operator product expansion. Throughout, we use the
kinematics defined in \cite{lastpap}. We first define the singlet quark
distributions,
\begin{equation}
K_{i\Sigma} (x, y) = \sum_a K_{ia}(x,y)
\end{equation}
where the sum is over all quark flavors. Their one-loop coefficient
functions are
\begin{equation}
C_{1,2\Sigma}^{(0)} \left( {x_B\over x}, {x_B\over y} \right)
= \bar e^2 y \delta( x - x_B ) \ ,
\end{equation}
where $\bar e^2 = \sum_i e_i^2/N_f$ is the mean square quark charge and
$N_f$ is the number of active quark flavors. Using the relation
\begin{equation}
\Delta q_T (x) = \frac{2}{x} \int dy
\left\{ K_1 (x, y) + K_2 (x, y) \right\} \ ,
\end{equation}
it is easy to see that the result is consistent with Eq.\ (1).

\begin{figure}[t]
\begin{center}
\vspace{0.5cm}
\mbox{
\begin{picture}(0,50)(100,0)
\put(-15,-14){\insertfig{8}{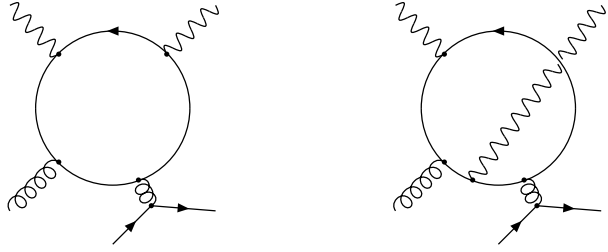}}
\end{picture}
}
\end{center}
\caption{\label{fig1} Feynman diagrams contributing to two quark and one
gluon intermediate states. The intermediate longitudinal gluon is onshell,
so a special propagator derived from the equation of motion has been
employed. There are a total of six diagrams. In dimensional regularization,
the sum is zero.}
\end{figure}

The general strategy of higher-twist factorization beyond the leading
orders has been presented in \cite{lastpap} and will not be repeated
here. For the singlet factorization, we need to consider four classes of
parton intermediate states: two quarks, two quarks-one gluon, two gluons,
and  three gluons. In the cases of two-quark and two-quark-one-gluon
states, many of the relevant diagrams have been considered in
\cite{lastpap}; we will not repeat that analysis here. In addition we must
take into account the explicit singlet diagrams shown in Fig.\ 1. A
detailed calculation shows that their contributions cancel. Thus the
one-loop coefficient functions of $K_{i\Sigma} (x, y)$ remain the same
as those of the non-singlet sector. In what follows, we focus entirely
on two- and three-gluon intermediate states.

The diagrams involving two-gluon intermediate states are shown in Fig.\ 2.
To isolate the ${\cal O}(1/Q)$ contribution, the incoming gluon is given
a transverse momentum which is expanded to first order in the internal
propagators. Our calculation shows that only the correlation function
\begin{equation}
{\mit\Gamma}_{2gB}^\mu (x, y) = \delta(x - y)
\int {d\lambda\over 2\pi} {\rm e}^{i\lambda x}
\langle PS |
A^\mu_a (0) i \partial_\alpha A^\alpha_a(\lambda n)
| PS \rangle \
\end{equation}
contributes, where $\alpha$ and $\mu$ take only transverse values.

\begin{figure}[t]
\begin{center}
\vspace{0.5cm}
\mbox{
\begin{picture}(0,50)(100,0)
\put(-90,-14){\insertfig{13}{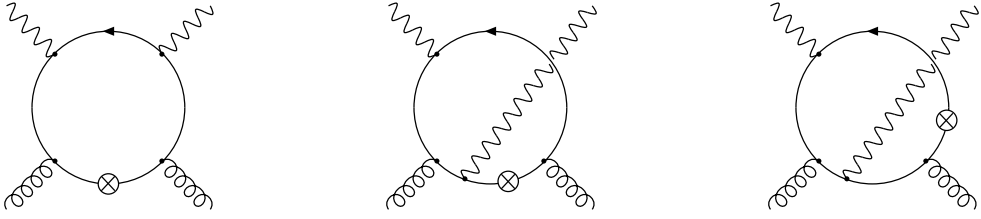}}
\end{picture}
}
\end{center}
\caption{\label{fig2} Two-gluon contributions to $g_T$. The cross
$\otimes$ represents one transverse momentum operator insertion. There
are a total of eight diagrams.}
\end{figure}

The perturbative diagrams corresponding to three gluon intermediate
states are shown in Fig.\ 3. The relevant correlation function is
\begin{equation}
{\mit\Gamma}_{3gB}^\mu (x, y)
= \int {d\lambda\over 2\pi} {d\mu\over 2\pi} {\rm e}^{i\lambda x}
{\rm e}^{i\mu (y - x)}
\langle PS|
( -i) {\sl g}_B f_{abc}
A^\mu_a (0) A_{b\, \alpha}(\mu n) A^\alpha_c (\lambda n)
|PS \rangle\ .
\end{equation}
All fields and couplings in the above expressions are bare. Gauge
invariance demands that the final result depends on the combination
\begin{eqnarray}
{\mit\Gamma}_{gB}^{\mu} (x, y)
&=&
{\mit\Gamma}_{2gB}^\mu (x, y) + {\mit\Gamma}_{3gB}^\mu(x, y)
\nonumber \\
&=& \int {d\lambda\over 2\pi} {d\mu\over 2\pi} {\rm e}^{i\lambda x}
{\rm e}^{i\mu(y-x)}
\langle PS |
A^\mu_a(0) i D_{\alpha a b} (\mu n) A^\alpha_b (\lambda n)
| PS \rangle
\end{eqnarray}
of distributions,
where $D_{\alpha ab}^\mu = \delta_{ab} \partial^\mu - {\sl g}
f_{acb} A_c$. In addition, gauge invariance demands that physical
observables depend on moments of
\begin{eqnarray}
xy {\mit\Gamma}_{gB}^{\mu} (x, y)
&=& \int {d\lambda\over 2\pi} {d\mu\over 2\pi}
{\rm e}^{i\lambda x} {\rm e}^{i\mu (y - x)}
\langle PS|
F^{+\mu}_a (0) iD_{\alpha a b} (\mu n) F^{+\alpha}_b (\lambda n)
| PS \rangle \nonumber \\
&=& K_{gB} (x, y) i \epsilon^{\alpha\beta\gamma\mu}
p_\alpha n_\beta S_\gamma \ .
\end{eqnarray}
With appropriate insertions of light-cone gauge links, which can be
generated by summing over states with additional longitudinally-polarized
gluons, $K_{gB}$ is gauge invariant. Our goal is to calculate its
one-loop coefficient function.

\begin{figure}[t]
\begin{center}
\hspace{0cm}
\mbox{
\begin{picture}(0,100)(100,0)
\put(-10,-14){\insertfig{8}{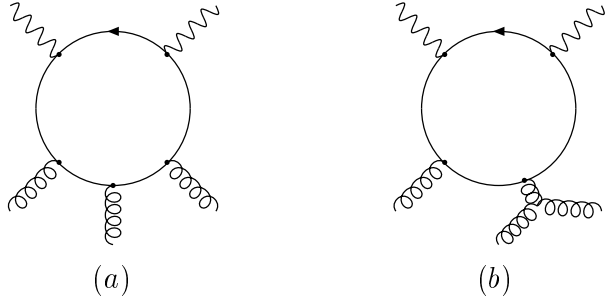}}
\end{picture}
}
\end{center}
\caption{\label{fig3} Feynman diagrams for three-gluon Compton scattering.
There are a total of 24 diagrams represented in (a). The other 8, in figure 
(b), have the form of those in Fig.\ 1, and vanish for the same reason.}
\end{figure}

The Compton amplitude for virtual photon scattering on a transversely
polarized nucleon can be written as ($\epsilon^{0123} = + 1$),
\begin{equation}
T^{\mu\nu}
= - i \epsilon^{\mu\nu\alpha\beta}q_\alpha S_{\perp\beta}
{1\over \nu} S_T (x_B, Q^2) \ .
\end{equation}
In QCD, we can write the singlet part as
\begin{eqnarray}
S_T^{\Sigma} (x_B, Q^2)
= \bar e^2 \int^1_{-1} \frac{dx}{x} \frac{dy}{y}
\left\lbrack\vphantom{M\left({xB \over y}\right)}\right.
\!\!\!&&\!\!\!\! M_{1\Sigma} \left( {x\over x_B}, {y\over x_B} \right)
K_{\Sigma 1B}(x, y)
\nonumber\\
\!\!\!&+&\!\!\! M_{2\Sigma} \left( {x\over x_B}, {y\over x_B} \right)
K_{\Sigma 2B} (x, y)
\\
\!\!\!&+&\!\!\!\left. \frac{N_f}{y - x}
M_{g} \left( {x\over x_B}, {y\over x_B} \right) K_{gB} (x, y) \right]
- (x_B \rightarrow - x_B) \ , \nonumber
\end{eqnarray}
where the $M$'s are perturbation series in $\alpha_s$ and have infrared
poles. Once again, the tree and one-loop level expressions for $M_{i\Sigma}$
are the same as the nonsinglet case in Ref. \cite{lastpap}.

The amplitude $M_g$ starts at the one-loop level. To simplify the
expression, we assume $|x_B| > 1$ so that the amplitude is purely real.

Since all Feynman diagrams are computed in bare perturbation theory, the
result depends on the bare coupling $g_B$. We replace it with the
renormalized coupling in the $\overline{\rm MS}$ scheme; the difference
appears only at higher orders in $\alpha_s (\mu^2)$. A straightforward
calculation yields
\begin{eqnarray}
M^{(1)}_g(x, y) &=& {\alpha_s(\mu^2)T_F\over 4\pi}
\left({\bar\mu^2\over Q^2}\right)^{\epsilon/2}
{1\over 2xy}\nonumber\\
&&\!\!\!\!\times\left\lbrace
\vphantom{{(3y^2-2x^2)(1-(x-y))+xy\over y-x}}
{2\over\epsilon}\left\lbrack
\left((y-2x)(1-x)+3y(y-x)-6\,{y\over x}\,(y-x)\right)
\log\left(1-x\right)\right.\right.\nonumber\\
&&+\left((4x-3y)(1-y)+4x(y-x)-8\,{x\over y}\,(y-x)\right)
\log\left(1-y\right)\nonumber\\
&&\left.+\left({(x+y)(2x-3y)\over x-y}(1-(x-y))-xy\right)
\log\left(1-(x-y)\right)\right\rbrack
\nonumber\\
&&-3\left(y+2\,{(y-x)^2\over x}\right)(1-x)\log(1-x)\nonumber\\
&&+\left( 
3(4x-3y)(1-y) + 14x(y-x) - 20\,{x\over y}\,(y-x) 
\right)\log(1-y) \nonumber\\
&&+{6x^2+xy-9y^2\over x-y}(1-(x-y))\log\left(1-(x-y)\right)\nonumber\\
&&-{1\over 2}\left((y-2x)(1-x)+3y(y-x)-6\,{y\over x}\,(y-x)\right)
\log^2\left(1-x\right)\nonumber\\
&&-{1\over 2}\left((4x-3y)(1-y)+4x(y-x)-8\,{x\over y}\,(y-x)\right)
\log^2\left(1-y\right)\nonumber\\
&&\left.-{1\over2}\left({(x+y)(2x-3y)\over x-y}(1-(x-y))-xy\right)
\log^2\left(1-(x-y)\right)\right\rbrace\;\; ,
\end{eqnarray}
where $\epsilon = 4-d$, $\bar\mu^2 = 4\pi e^{-\gamma_E}\mu^2$, and
$\gamma_E$ is the Euler constant.  We have also introduced the generator
normalization $T_F = 1/2$.

Now we want to show that $S_T^{\Sigma} (x_B, Q^2)$ is factorizable at
the one-loop level, i.e., the infrared poles $1/\epsilon$ in $M_g$ match
the ultraviolet poles in $K_{gB}$. To this end, we use the infrared
poles in $M_g$ to generate a scale evolution equation for the renormalized
parton distributions,
\begin{eqnarray}
{d\over d \log \mu^2}\!\!\!&&\!\!\!\int^1_{-1} dxdy\left({2\over
 x(x_B-x)}\right)(K_{1\Sigma}(x,y,\mu^2) + K_{2\Sigma}(x,y,\mu^2))
\nonumber\\
&=&
{\alpha_s(\mu^2) N_f T_F \over8\pi}
\int^1_{-1} {dxdy\over x^2y^2(x-y)}
\nonumber \\
&&\times \left\lbrack
\left((y-2x)(x_B-x)+3y(y-x)-6\,{x_By\over x}\,(y-x)\right)
\log\left(1-{x\over x_B}\right)\right.\nonumber\\
&&\quad+\left((4x-3y)(x_B-y)+4x(y-x)-8\,{x_Bx\over y}\,(y-x)\right)
\log\left(1-{y\over x_B}\right)\nonumber\\
&&\!\!\!\!\!\!\!\!\!\!
\left.+\left({(x+y)(2x-3y)\over x-y}\,(x_B-(x-y))-xy\right)
\log\left(1-{x-y\over x_B}\right)\right\rbrack
K_g(x,y) + \cdots\;\; ,
\label{evolve}
\end{eqnarray}
where the ellipses denote the homogeneous part of the evolution. This
contribution is given explicitly in Ref.\ \cite{lastpap}. In order to
compare this result to the known twist-three evolution \cite{bkl},
we must first remove the twist-two contribution. Consider the twist-two
operator
\begin{equation}
\theta^{\mu_1\mu_2\cdots \mu_{n+1}}_{n+1}
= F^{\alpha(\mu_1} iD^{\mu_2} \cdots iD^{\mu_n}
i\tilde F^{\mu_{n+1})}_{\;\;\;\alpha}\;\; ,
\end{equation}
whose indices have been symmetrized and whose traces have been removed.
Its matrix elements in the nucleon state are given by
\begin{equation}
\langle PS|\theta^{\mu_1\cdots \mu_{n+1}}_{n+1}|PS\rangle
= 2a_n P^{(\mu_1}\cdots P^{\mu_n} S^{\mu_{n+1})}\;\; .
\end{equation}
It is the $\perp+\cdots+$ component of $\theta$ that appears
in this transverse process :
\begin{eqnarray}
\theta_{n+1}^{(i+\cdots)}
= {2\over n+1}\epsilon^{ij}
\left\lbrack\vphantom{\sum_{m=0}^{n-1}}
(n+1) iA^j (i\partial^+)^n i\partial^k A^k
+ iA^j (i\partial^+)^{n-1} D_\alpha F^{\alpha +} \right.
\nonumber\\
+ \left. f^{abc} A_a^j (i\partial^+)^{n-1} {\sl g} A_b^k i\partial^+ A_c^k
+ \sum_{m=0}^{n-1}
f^{abc} A_a^j( i\partial^+)^m {\sl g} A_b^k (i\partial^+)^{n-m} A_c^k
\right\rbrack\;\; .
\label{tw2}
\end{eqnarray}
Here, latin indices are understood to take values in the transverse
dimensions. Expanding Eq.\ (\ref{evolve}) in the large $x_B$ limit,
one arrives at evolution equations for the moments of the parton
distributions. Removing the twist-two part of these equations via Eq.\
(\ref{tw2}), one can obtain the evolution of the twist-three operators.
A detailed check shows that our result is identical to that found in
Ref.\ \cite{bkl} obtained by studying the ultraviolet divergences present
in the twist-three operators. We note here that this separation provides
a new homogeneous term in the evolution of the singlet twist-three quark
operators. In the absence of this term, the diagonal evolution of these
operators would be identical to that in the nonsinglet sector since the
contributions displayed in Fig.\ 1 vanish.

The final step of the calculation is to take the imaginary part of the
factorized $S_T^{\Sigma}(x_B, Q^2)$ to get a factorized expression for
the structure function $g_T^{\Sigma} (x_B, Q^2)$ in the physical region
$x_B<1$. We find the coefficient function
\begin{eqnarray}
&&C_g^{(1)}\left({x_B\over x}, {x_B\over y}\right) =
{\alpha_s(\mu^2)T_F\over 8\pi}
{1\over 2xy}\left\lbrack
3\left(y+2\,{(y-x)^2\over x}\right)(x_B-x)\theta\left({x\over x_B}-1\right)
\right.\nonumber\\
&&-\left(3(4x-3y)(x_B-y) + 14x(y-x) - 20\,{x_B x\over y}\,(y-x)\right)
\theta\left({y\over x_B}-1\right)\nonumber\\
&&-{6x^2+xy-9y^2\over x-y}(x_B-(x-y))
\theta\left({x-y\over x_B}-1\right)\nonumber\\
&&+\left((y-2x)(x_B-x)+3y(y-x)-6\,{x_By\over x}\,(y-x)\right)
\log\left({x\over x_B}-1\right)
\theta\left({x\over x_B}-1\right)\nonumber\\
&&+\left((4x-3y)(x_B-y)+4x(y-x)-8\,{x_B x\over y}\,(y-x)\right)
\log\left({y\over x_B}-1\right)\theta\left({y\over x_B}-1\right)\nonumber\\
&&\left.+\left({(x+y)(2x-3y)\over x-y}(x_B-(x-y))-xy\right)
\log\left({x-y\over x_B}-1\right)\theta\left({x-y\over x_B}-1\right)
\right\rbrack\;\; ,
\end{eqnarray}
where $\theta(x)$ is the step-function, which appears in Eq.\ \ref{resuts} along
with a factor of $1/(y-x)$.

To check the Burkhardt-Cottingham sum rule \cite{bc}, we integrate
$g_T^{\Sigma}(x_B,Q^2)$ over $x_B$.  Assuming the integration over
$x$ and $y$ can be interchanged with that of $x_B$, one obtains
\begin{equation}
\int^1_0 dx_B g^{\Sigma}_T(x_B, Q^2)
= {1\over 2} \bar e^2
\left(1-{7\over 2}C_F{\alpha_s(Q^2)\over 2\pi}\right)
\langle PS | \sum_i \bar\psi_i \gamma_\perp \gamma_5 \psi_i
| PS \rangle \ ,
\end{equation}
where $\gamma_5$ has been defined in the 't Hooft-Veltman scheme. The
coefficient $7/2$ reduces to $3/2$ if we define $\gamma_5$ so that the
nonsinglet axial current is conserved. Compared with the factorization
formula for $g_1(x_B, Q^2)$ \cite{kodaira}, we have the Burkhardt-Cottingham
sum rule at one loop;
\begin{equation}
\int^1_0 dx_B g_2^{\Sigma} (x_B, Q^2) = 0 \ .
\end{equation}
If the order of integration cannot be interchanged because of the singular
behavior of the parton distributions at small $x$ and $y$, the above sum
rule may be violated. Indeed, some small $x_B$ studies indicate such
singular behavior \cite{smallx}.

Finally, we consider the next-to-leading order correction to the singlet
part of the $x^2$ moment of $g_T (x_B, Q^2)$. In the leading order, it is
well known:
\begin{equation}
\int^1_0 dx \ x^2 g_T^{\Sigma} (x, Q^2)
= {1\over 3} \bar e^2
\left( {1\over 2}a_{2\Sigma}(Q^2) + d_{2\Sigma}(Q^2) \right) \ ,
\end{equation}
where $a_2$ is the second moment of the $g_1(x,Q^2)$ structure
function and $d_2$ is a twist-three matrix element \cite{jj}.
Using the coefficient functions found above, we have
\begin{eqnarray}
\int^1_0 dx x^2 g_T^{\Sigma} (x, Q^2)
&=& {1\over 3} \bar e^2
\left\lbrace{a_{2\Sigma}(Q^2)\over 2}
\left\lbrack 1 + {\alpha_s(Q^2)\over 4\pi} {7\over 12} C_F \right\rbrack
- {a_{2g}(Q^2)\over 2} \; {\alpha_s(Q^2)\over 4\pi} \;
{5\over3} N_f T_F \right. \nonumber \\
&&\phantom{{1\over 3}\sum_i\hat e_i^2}\quad
\left. + d_{2\Sigma}(Q^2)\left\lbrack1
+ {\alpha_s(Q^2)\over 4\pi}\left({27\over 4}C_A
- {29\over 3}C_F + {10\over3} N_f T_F \right)
\right\rbrack\right\rbrace\ .
\end{eqnarray}
Using the next-to-leading result for $g_{1\Sigma}(x, Q^2)$ \cite{kodaira},
we find
\begin{eqnarray}
\int^1_0 dx x^2 \!\!\!&&\!\!\! \left(
g^{\Sigma}_T (x, Q^2) - {1\over 3} g^\Sigma_1(x, Q^2)
\right) \nonumber\\
&& = {1\over 3}\bar e^2
d_{2\Sigma}(Q^2) \left\{ 1 + {\alpha_s(Q^2)\over 4\pi}
\left(
{27\over 4} C_A - {29\over 3} C_F + {10\over3} N_f T_F
\right) \right\} \ .
\end{eqnarray}
Notice that the combination of $g_T$ and $g_1$ relevant to $d_{2\Sigma}$
receives no radiative correction, as in the nonsinglet sector. A detailed
calculation shows that analogous results are valid for all higher moments
as well \cite{newpap}. This implies that the tree relation between
$g_1(x,Q^2)$ and the twist-2 part of $g_2(x,Q^2)$ is respected at one-loop
order. These results and their implications will be presented in a future
communication \cite{newpap}.

The significance of the present result is as follows. In the leading order
analysis of $g_2$, one just needs the leading-logarithmic evolution of
$K_i (x, y)$ which is now well known \cite{bkl,Rat,evolve1}, including its
large $N_c$ behavior \cite{largenc}. In the next-to-leading order, one
needs to know the coefficient functions and the two-loop evolution of
$K_i (x, y)$. The former is now complete with Ref.\ \cite{lastpap} and the
present paper. The latter has not yet been calculated, but in generally
its effort is not as important as the coefficient function we calculate here.

Finally, let us add a few remarks on the comparison of our findings with a
previous calculation of Ref.\ \cite{Bel}. As has been noted earlier
\cite{Bel00} the result of that paper is not complete since the
contributions of twist-three two-gluon operators, i.e.\ the diagrams on
Fig.\ \ref{fig2}, were not accounted for. Since the effect of two-gluon
graphs on the final answer is reduced to a redefinition of the correlation
function (schematically) from $\langle A^\perp_\mu A^\perp_\nu A^\perp_\rho
\rangle$ to $\langle A^\perp_\mu D^\perp_\nu A^\perp_\rho \rangle$ without
affecting the three-gluon coefficient function, we can directly compare both
results. Before we do it, we observe that the basis of three-gluon functions
used in \cite{jiold,Bel} is redundant, namely, in the decomposition
\begin{equation}
\langle A_\mu A_\nu A_\rho \rangle
\propto N (x_1, x_2)
g^\perp_{\mu\rho} \epsilon_{\nu - + \sigma} s^\perp_\sigma
+ \widetilde N (x_1, x_2) s^\perp_\rho \epsilon_{\mu \nu - +}
+ \dots ,
\end{equation}
obviously, the Lorentz structure in front of $\widetilde N$ can be
expressed in terms of the first line using an identity
\begin{equation}
s^\perp_\rho \epsilon_{\mu \nu - +}
= g^\perp_{\rho \nu} \epsilon_{\mu - + \sigma} s^\perp_\sigma
- g^\perp_{\rho \mu} \epsilon_{\nu - + \sigma} s^\perp_\sigma .
\end{equation}
Therefore, in final equations of \cite{Bel} we have to put $\widetilde N
= 0$ to be consistent with the present analysis. Moreover, the result in
Ref.\ \cite{jiold} must be also modified correspondingly.

Next, to have a correspondence with the correlation functions used here
and in Ref.\ \cite{Bel}, we have identify the momentum fractions of the
gluon lines as follows: $x = x_1$ and $y = x_1 - x_2$. For the latter
convenience we introduce as well a new coefficient function
\begin{equation}
{\cal E}^{(1)}_g (x, y; x_B) \equiv \frac{1}{y - x} C^{(1)}_g
\left( \frac{x_B}{x}, \frac{x_B}{y} \right) .
\end{equation}
Since the function $N (x_1, x_2)$ is symmetric w.r.t.\ interchange
of its arguments only symmetric part of ${\cal E}^{(1)}_g (x_1, x_1 - x_2;
x_B)$ will survive. Thus, a symmetrization leads to the equation
\begin{equation}
{\cal E}^{(1)}_g (x_1, x_1 - x_2; x_B)
+ {\cal E}^{(1)}_g (x_2, x_2 - x_1; x_B)
= \frac{\alpha_s}{8 \pi} E_2 (x_1, x_2; x_B) ,
\end{equation}
where $E_2$ is the coefficient function from Ref.\ \cite{Bel}.

\vspace{0.5cm}

The authors would like to thank V.M. Braun and G.P. Korchemsky for useful
discussions on the subject of this paper. In addition, we wish to
acknowledge the support of the U.S. Department of Energy under grant no.
DE-FG02-93ER-40762.


\begin{thebibliography}{99}
\bibitem{hey}
A.J.G. Hey, J.E. Mandula, Phys. Rev. D 5 (1972) 2610;\\
M.A. Ahmed, G.G. Ross, Nucl. Phys. B 111 (1976) 441;\\
K. Sasaki, Prog. Theor. Phys. 54 (1975) 1816.
\bibitem{exp}
SMC Collaboration, D. Adams et al., Phys. Lett. B 336 (1994) 125;\\
E142 Collaboration, P.L. Anthony et al., Phys. Rev. Lett. 71 (1993) 959;
Phys. Rev. D 54 (1996) 6620;\\
E143 Collaboration, K. Abe et al., Phys. Rev. Lett. 76 (1996) 587;
Phys. Rev. D 58 (1998) 11203;\\
E154 Collaboration, K. Abe et al., Phys. Lett. B 404 (1997) 377;\\
E155 Collaboration, P.L. Anthony et al., Phys. Lett. B 458 (1999) 530.
\bibitem{Shur}
E.V. Shuryak, A.I. Vainshtein, Nucl. Phys. B 199 (1982) 951.
\bibitem{bkl}
A.P. Bukhvostov, E.A. Kuraev, L.N. Lipatov, JETP Lett. 37 (1983) 484;
Sov. Phys. JETP 60 (1984) 22.
\bibitem{Rat}
P.G. Ratcliffe, Nucl. Phys. B 264 (1986) 493.
\bibitem{lastpap}
X. Ji, W. Lu, J. Osborne, X. Song, hep-ph/0006121, Phys. Rev. D (in
press).
\bibitem{Bel}
A.V. Belitsky, A.V. Efremov, O.V. Teryaev, Phys. Atom. Nucl. 58 (1995) 1253.
\bibitem{bc}
H. Burkhardt, W.N. Cottingham, Ann. Phys. 56 (1970) 453.
\bibitem{kodaira}
W.A. Bardeen, A.J. Buras, D.W. Duke, T. Muta, Phys. Rev. D 18 (1978) 3998;\\
J. Kodaira, S. Matsuda, K. Sasaki, T. Uematsu, Nucl. Phys. B 159 (1979) 99.
\bibitem{smallx}
I.P. Ivanov, N.N. Nikolaev, A.V. Pronyaev, W. Sch\"afer, Phys. Lett.
B 457 (1999) 218.
\bibitem{jj}
R.L. Jaffe, X. Ji, Phys. Rev. D 43 (1991) 724.
\bibitem{newpap}
X. Ji, J. Osborne, to be published.
\bibitem{evolve1}
I.I. Balitsky, V. M. Braun, Nucl. Phys. B 311 (1989) 541;\\
X. Ji, C. Chou, Phys. Rev. D 42 (1990) 3637;\\
B. Geyer, D. M\"uller, D. Robaschik, hep-ph/9611452;\\
J. Kodaira, Y. Yasui, K. Tanaka, T. Uematsu, Phys. Lett. B 387 (1996) 855.
\bibitem{largenc}
A. Ali, V. M. Braun, G. Hiller, Phys. Lett. B 266 (1991) 117;\\
X. Ji, J. Osborne, Eur. Phys. J. C 9 (1999) 487;\\
A.V. Belitsky, Phys. Lett. B 453 (1999) 59; Nucl. Phys. B 558 (1999) 259;\\
S.E. Derkachov, G.P. Korchemsky, A.N. Manashov, Nucl. Phys. B 566 (2000)
203;\\
V.M. Braun, G.P. Korchemsky, A.N. Manashov Phys. Lett. B 476 (2000) 455.
\bibitem{Bel00}
A.V. Belitsky, Nucl. Phys. B 574 (2000) 407.
\bibitem{jiold}
X. Ji, Phys. Lett. B 289 (1992) 137.
\end{thebibliography}
\end{document}